# Radial segregation driven by axial convection

A. C. Santomaso, L. Petenò and P. Canu[1]
*DIPIC - University of Padua, via Marzolo 9, 35131 Padua, Italy.*

**Abstract.** - We experimentally study the mixing of binary granular systems in a horizontal rotating cylinder. When materials the have same size and differ by dynamic angle of repose only, we observe an axial transport of matter that generates transient radial segregation. The system then evolves towards homogeneity. If materials differ by density also radial segregation becomes steady. A mechanism is suggested where radial segregation is promoted by axial differences of dynamic angle of repose. This differs from the free surface segregation suggested so far to explain radial segregation.

*Introduction.* - Mixing of granular materials is a common industrial operation with many interesting speculative issues. It is well known that particles in horizontal cylinders with different size or density may give segregation radially on short time scales and axially on longer time scales [1-4]. However other variables such as friction [4], particle shape [5], dynamic angle of repose [3, 6], can influence mixing dynamics. A satisfactory explanation encompassing all the observed cases is lacking yet. Traditionally, the dynamics of the horizontal cylinder at low rotational speed ($\Gamma = \omega^2 R/g$ ranging from $10^{-4}$ to $10^{-2}$, with $\omega$ rotational speed [rad/s], $R$ cylinder radius [m] and $g$ gravitational acceleration [m/s$^2$]) is described with a 2D model, based on kinematics of the particles in the transverse planes. Two regions are identified: a thick bulk of material rotating with the cylinder as a solid body until it reaches its dynamic angle of repose $\delta$, and a free flowing region confined to the thin surface layer [7]. The interface between the two regions is characterized by a continuous mass transfer from the bulk to the surface layer in its upper part and vice versa in the lower one. Mixing mainly occurs in the thin surface layer being particle collisions and percolation negligible in the thick bulk [7]. Also segregation is affected by the surface dynamics. Axial segregation is associated to local axial gradient of the surface layer [2, 6], while for radial segregation, the most accredited mechanism [5] assumes the surface layer to influence the inner composition, because of its sieving effect which enables selective percolation of the denser and/or smaller particles towards the core of the bed [8]. In the case of 3D systems initially segregated side by side, numerical simulations [9] and experimental findings [10] suggest that the dynamics leading to radial segregation can be much richer, possibly connected to axial displacements of material. Considering the interplay between particle size and density, two distinct behaviours of the system have been observed by simulation [9]. When the smaller particles are heavier, the system goes quickly towards a radially segregated and stable state. Segregation however can be counterbalanced by making the smaller particle lighter. The system then develops an unstable core resulting in the propagation of a gradually dissolving "segregation wave" [9]. The process leads to a homogeneous mixture. Such behaviour has not been confirmed by experiments so far, to our knowledge. In the present experimental work, we studied a phenomenon that bears some analogies with that presented above because of the appearance of a segregation wave. Here however not the particle size, but $\delta$ (i.e. the slope of the free surface) differentiates the materials. The interplay between $\delta$ and particle density is investigated as well. We observed that convective axial displacements of material are generated on the surface whenever small differences of $\delta$ arise along the cylinder [10].

---

[1] E-mail: paolo.canu@unipd.it

Table 1. - Granular materials and combinations with resulting mixture features.

| Material | Colour | $d$ [μm] | Roundness | $\delta$ [°] | $\rho$ [kg/m$^3$] | Combination | $\Delta$ [°] | $\rho_W/\rho_B$ |
|---|---|---|---|---|---|---|---|---|
| I - taed | Blue | 300-500 | 0.75 | 34.0 ± 0.3 | 1330 | A: I+II | 1.0° | 1 |
| II - taed | White | 300-500 | 0.73 | 35.0 ± 0.3 | 1330 | | | |
| III - taed | Blue | 500-600 | 0.70 | 37.7 ± 0.2 | 1330 | B: III+IV | 1.9° | ~1.5 |
| IV - sp | White | 500-600 | 0.82 | 35.8 ± 0.3 | 1990 | | | |
| V - sp | White | 500-600 | 0.87 | 32.7 ± 0.1 | 1990 | C: III+V | 5.0° | ~1.5 |

So we suggest a mechanism where a convective axial flow below the surface arises in order to counter-balance the surface flow, generating the observed segregation patterns. In particular if particles have the same density, these flows generate a segregation wave that similarly to [9] leads to a well-mixed system. If differences in density are present also the wave leads to stable radial segregation along the cylinder.

*Experimental set-up* - Different binary combinations of granular materials have been studied. In each combination, particle size is comparable but $\delta$, particle density and colour (so to distinguish the materials) differ. Combinations and components features are summarized in Tab. 1. We always mixed some blue tetraacetylenethyldiammine (taed) powder with a white material to allow composition quantification by image analysis. Sodium percarbonate (sp) powders differ by shape, material *IV* is spheroidal while *V* is irregular, resulting in quite a large difference of $\delta$, $\Delta=|\delta_W-\delta_B|$. $\Delta$ is consistent with microscopically observed differences of particle roundness ($4\pi$Area/Perimeter$^2$). From Tab.1 roundness follows this order: white taed<blue taed<irregular sp<spheroidal sp. A deeper study on the relationship between particle shape and $\delta$ and its implications in mixing is underway and will be reported elsewhere. We used a horizontal cylinder, 0.11 m long with $R$=0.046 m. The cylinder was always filled up to 30% in volume with two different materials (volume ratio 1:1) arranged side by side i.e. completely axially segregated. To study the 3D composition map, a solidification technique of the mixture was used [10]. After a fixed number of revolutions at low rotational speed ($\Gamma$ =1.84·10$^{-3}$) the mixture was saturated with molten wax, then solidified and cut in nine slices (fig. 1). Both surfaces of each slice were digitized, stored as a grey scale image (8 bits) and analyzed obtaining 18 radial maps of concentration along the axis. Using this technique it is not possible to resume the experiment after solidification and this represent its major limitation. Such a limitation is compensated by the detailed information about the radial composition distribution in the slicing plane, by the ease of application, by the possibility of preserving the samples indefinitely.

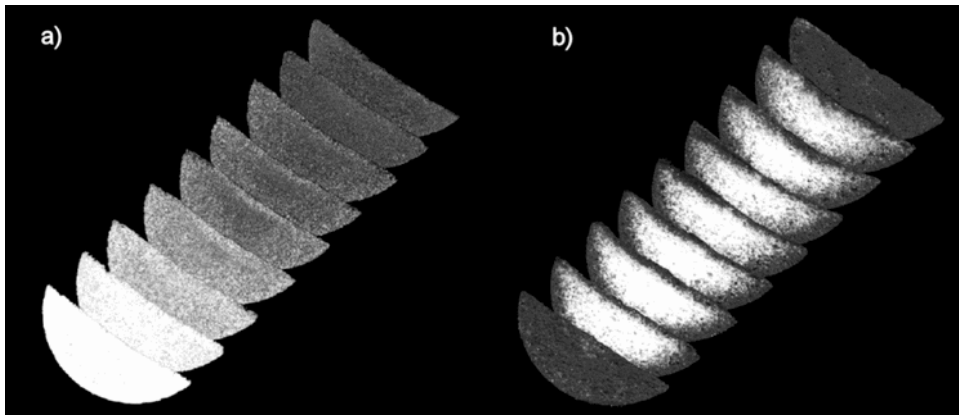

Fig. 1. - Solidified bed of blue and white taed (combination *A*) (a) and bed of blue taed and spheroidal percarbonate (combination *C*) (b), after 200 revs.

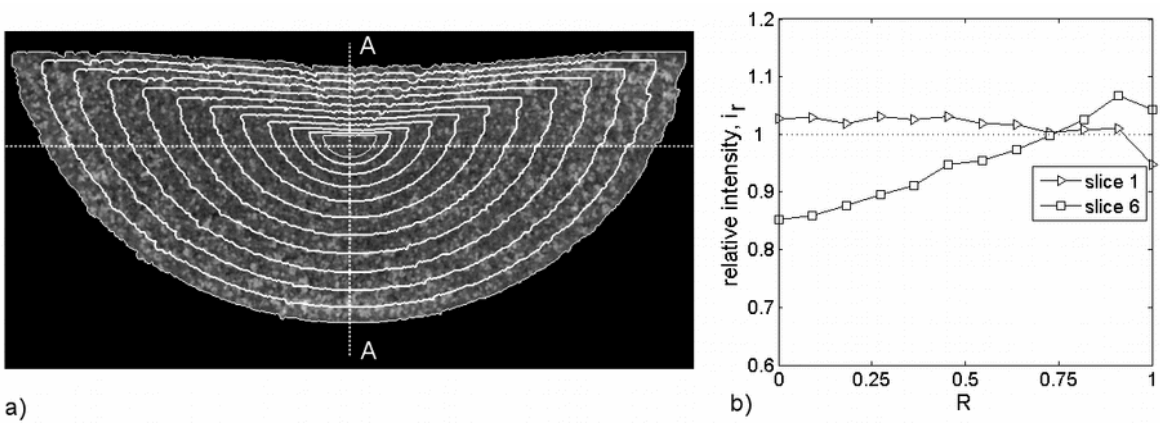

Fig. 2. - Sample image of a slice surface (combination A, after 200 revs., slice 6) (a) and analysis of concentration along concentric rings (b); R=0 is the central ring.

The 3D composition field after solidification can also be reconstructed, similarly to more sophisticated techniques such as NMR [3, 8, 11] and PEPT [12], which provide better time resolution at the expenses of spatial details. As a compromise between destructive nature of the technique and reproducibility each experiment was performed twice. Replicated experiments confirmed that an accurate procedure resulted in reproducible data. Image analysis of the slices was based on grey intensity, which corresponds to the local concentration of coloured particles. It yields a complete, continuous composition map $C(x,y)$ at each surface, for 10 axial positions, so that the complete $C(x,y,z)$ can be reconstructed, although $z$ does not vary continuously, and only some time instants are available. It is quite a large amount of information that requires some synthesis to formulate quantitative comparisons among different experiments. Consider a sample slice as shown in fig. 2a. To investigate the radial composition and quantitatively discuss about radial segregation we defined a conventional centre, approximately the centroid of the segregated core, and we identified 12 rings, paralleling the slice contour and approximating isoconcentration regions. (see fig. 1). We calculated the relative grey intensity, $i_r$, on each ring, defined as the mean value within the ring divided by the average intensity of the whole surface.

A plot of ring relative intensity vs. radial distance from the conventional centre is shown in fig. 2b. A truly homogeneous surface should result in a horizontal line at relative intensity equal to 1. Departures from the line $i_r=1$ quantify radial composition non-uniformity. With Fig. 2b in mind, we can measure a scaled deviation from uniformity as $s=|s^*-n_s|$, where $s^*=1-\min(i_r)/\max(i_r)$ and $n_s=(s_1^*+s_9^*)/2$. The index $s^*$ applies to each surface. It synthetically quantifies radial segregation at a given axial position. However, even without segregation $\max(i_r)$ and $\min(i_r)$ are never exactly the same. Because of the experimental technique, uneven binder distribution can generate a small disturbance, $n_s$. It can be evaluated by $s^*$ at the two extremities where there is no segregation, so $s^*$ just measures experimental noise. The average value of noise at the extremities has been subtracted to $s^*$ obtaining the scaled index $s$, allowing for comparison of different experiments, with different materials. s is approximately 0 for uniform surfaces and approaches 1 for fully segregated ones.

*System without density differences.* - Differently from many literature studies on segregation [1-6] where the particle size ratio is large, in case A size ratio is approx 1:1, and densities are precisely the same. In our experiments [10] we observed that the difference of $\delta$ (i.e. of free surface slope) between the two regions (blue and white) caused an axial convective flow on the free surface as sketched in fig. 3a. The material with the larger $\delta$ (1 in fig. 3a) rises the upper edge of the surface layer higher than the other material (2 in fig. 3a). Consequently, some axial drifting of material 1 above 2 takes place because of gravity. The process is more active at the beginning when pure materials are facing and it has a preferential direction from region 1 to region 2, the lower edges being closer in height. In order to counterbalance such surface flow, and to preserve the mass balance in the cylinder (i.e. to keep a constant level), an axial flow with opposite direction must develop somewhere. We suggest that this counter-flow takes place below the free surface. The possibility of axial displacement below the surface is not intuitive and rarely suggested in the literature as a possible mechanism of solid mixing. Considerations on particle mobility in deep layers [13], experimental results [10] and computer simulations [9] however suggest that close to the core slow rearrangements are indeed likely. In particular computer simulations [9] have shown that mixing particles with different size and density, larger and denser particles (density ratio 1:2) segregated in the core. However they were never observed close to the free surface, strongly indicating that the propagation mechanism was not due to surface flow but rather to a pure core flow. On the other hand, the classical segregation mechanism, i.e. percolation, based on difference of particle size and density is here inappropriate being size and density the same. The proposed segregation mechanism promotes also different radial mixing in the two halves of the cylinder. Material 1 rapidly mixes with material 2 when flowing on the surface in side 2, creating here a peripheral region enriched of material 1. The amount of 2 that slowly penetrate side 1 below the surface, moves in a region where chances of mixing are dramatically reduced. Particles are more constrained here and consequently material 2 remains concentrated creating the segregated core. The radial segregation starts from the interface and slowly extends towards the two ends of the cylinder. However axial motion of material 1 on the thin surface of region 2 is faster than displacement of material 2 below the surface in region 1; so the two segregation fronts move in opposite directions with different velocity. The asymmetric segregation after 200 revs. can be quantified by index *s* as shown in fig. 3b. Arrows indicate two axial positions (slices 3 and 7) equidistant from the initial interface (slice 5). Here segregation is larger on the side where material 1 flowed

(axially) faster i.e. on the surface of material 2. Results indicate also that segregation fades out at longer mixing time. Radial segregation is therefore a transient phenomenon for materials I and II, with two propagation fronts that can be described as a wave starting from the initial interface and spreading along the cylinder towards the ends. The extinction of the segregated core can be visualized also using the radial composition profiles (figs. 4a,b). At 1600 revs. all the radial profiles cluster to the $i_r=1$ line indicating uniform radial composition throughout the whole cylinder. The simple mechanism suggested above explains also the transient nature of the segregated core. As mixing proceeds, the driving force of axial displacements $\Delta$ gradually vanishes. Normal processes of radial mixing (by continuous mass transfer between free surface and core) and symmetric axial dispersion on the surface instead keep operating, ultimately leading to complete homogenization.

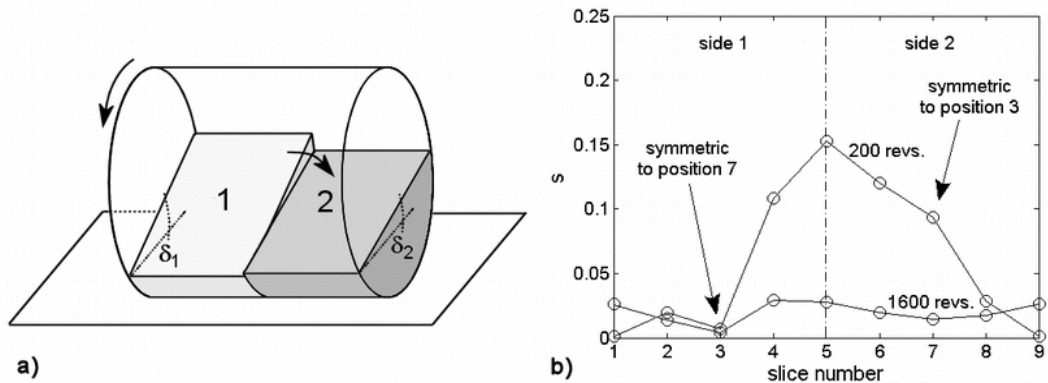

Fig. 3. - Differences of $\delta$ cause surface levels to be unbalanced, driving the axial convective flows (a). Radial segregation index of system A for 200 and 1600 revs. (b).

The segregated core is therefore progressively eroded and material is redistributed both radially and axially. These results show analogies with numerical DEM simulations [9] because of the transient nature of the segregation waves. In that case, however, it was the combination of different sizes and density ratios to drive the phenomenon. We suggest the possibility that differences of surface slope because of $\delta$, not reported by the Authors [9], could have played a role in that case also being $\delta$ a complex function of many variables including size and density.

*Effects of density differences*. - A different result is obtained if powders differ not only by $\delta$ but also by density. The sliced bed for combination *C* is shown in fig. 1b. Density ratio was about 1:1.5, while size ratio was again 1:1. Comparison with fig.1a (case *A*) indicates now a much stronger radial and axial segregation. Here the core is white material because it has now lower $\delta$. Note that blue material (higher $\delta$), completely envelops the white one. Quantitative information are shown in figs. 4c,d comparing also combination *B* and *C* at different times. In both cases large departure from uniform radial composition can be noticed for all slices excepted the extremities. These are made of almost pure material 1 and are actually two bands, just few particle diameters thick. Differently from case *A,* radial segregation in case *B* and *C* is permanent and the bed evolves faster to its final state, because $\delta$ is higher. Differences can be noticed between cases *B* ($\Delta=1.9°$) and *C* ($\Delta=5°$) also. When $\Delta$ is smaller segregation is definitely less intense. This is shown in fig. 5 that compares radial segregation along the cylinder for all cases together. Combination *C* reaches the largest values of radial segregation intensity along the whole cylinder excepted for the ends that are almost uniform material 1.

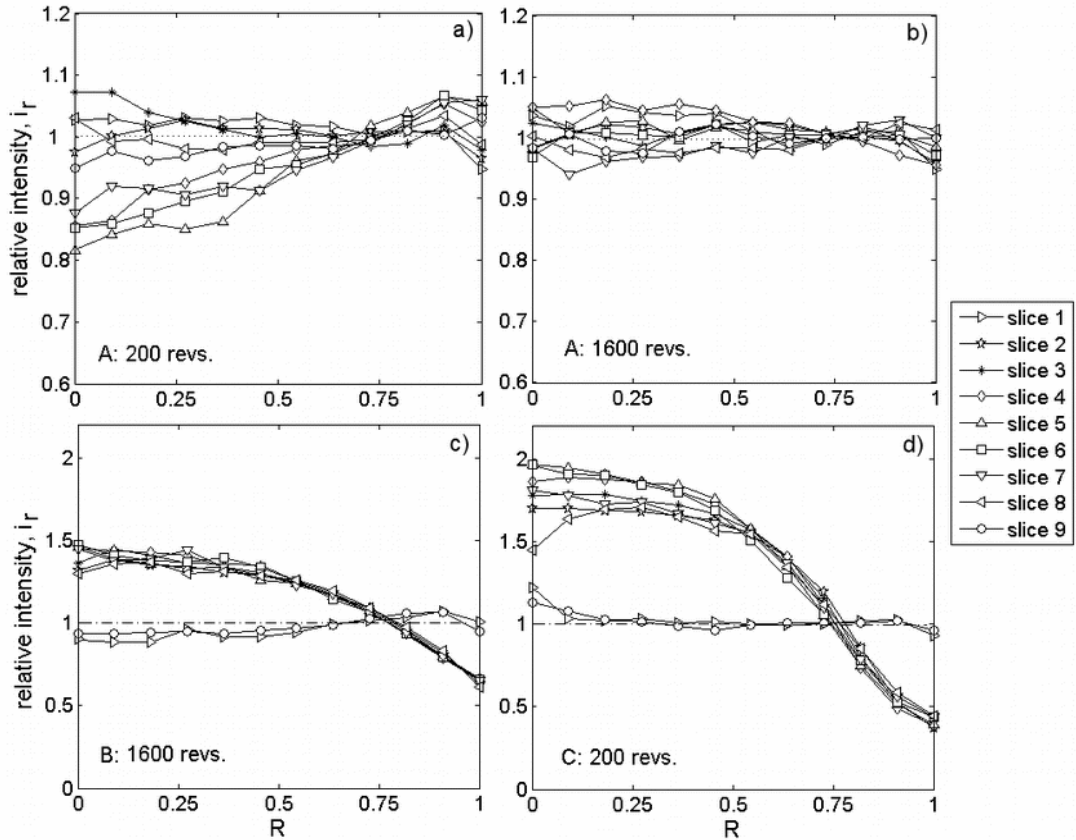

Fig. 4. - Radial segregation through relative grey intensity on concentric rings. Case A, after 200 revs. (a) and 1600 revs. (b). Case *B* after 1600 revs. (c), and case *C* after 200 revs. (d).

Also kinetics is affected by $\Delta$. Combination *C* reaches a steady segregation state in less then 200 revs. while combination *B* after the same time is still developing the final pattern, and stabilizes at 1600 revs. An experiment after 10000 revs. gave indeed the same axial distribution of radial segregation measured after 1600 revs. Moreover, combination *B* while developing its final configuration shows an asymmetric profile, similarly to combination *A*. After the segregation front dissolves however, mixture *B* remains steadily segregated. In addition to the mechanism suggested above (material 1 with higher $\delta$ that initially flows on the surface of material 2 inducing inner flow of material 2 to side 1) now material 1 is also lighter than 2. Because of the density ratio, mixing with material 2 is difficult in the surface layer. The lighter material 1 simply floats over the heavier one, thus remaining always in the periphery. In side 2 the radial segregation is therefore sharp as can be noticed by the peak in the profile at 200 revs. Moreover in case *B* notwithstanding the same density ratio $\rho_W/\rho_B$ as in case *C* segregation is less intense possibly because of a reduced difference of $\delta$. Looks like $\Delta$ plays a role in controlling the kinetics and the extent of segregation while $\rho_W/\rho_B$ decides if segregation is permanent or not. However experiments were limited to small $\Delta$, when $\rho_W/\rho_B=1$ (case *A*), and we can not exclude that large $\Delta$ with comparable density may cause a permanent radial segregation. This is however unlike because during mixing $\Delta$ will vanish, removing the 'driving forces' for radial segregation. In [9] it was proved that segregation can be avoided by balancing the opposite effects of size and density. We therefore suggest that something similar could happen for $\Delta$ and density. In our combinations $B$ and $C$ the material with higher $\delta$ was also the lighter one so it moved from side 1 to side 2 floating on the surface. What would happen if this material was the heavier one? There probably would be some percolative radial migration from the surface toward the core resulting in some degree of mixing. Unfortunately such experimental investigations are complicated by the difficulty of finding materials that exactly match the required properties. Hopefully, DEM simulations can help to confirm or reject such hypothesis, since they proved able to reproduce experimental $\delta$ values [14].

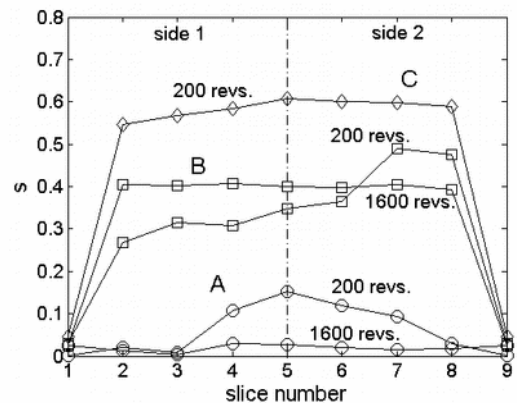

Fig. 5. - Comparison of axial distribution of $s$ for all the cases studied. In the starting configuration the material with higher $\delta$ was always on side 1, i.e. white for combination $A$ and blue for combinations $B$ and $C$.

*Conclusions.* - Radial segregation in partially filled, rotating cylinders has been studied through a solidification technique, coupled with digital image analysis. Binary systems with same size in combinations with differences of density and of dynamic angle of repose have been studied. A mechanism based on differences of dynamic angle of repose was suggested where two axial convective flows with opposite directions develop both on the surface layer and below it. Competition exists between these flows (leading to axial homogeneity) and the transverse displacement of particles (leading to radial homogeneity). This generates two segregation fronts that move from the centre towards the extremities and progressively dissolve leaving the system segregated or homogeneous according to the existence or not of density differences. Both intensity and kinetics of the segregation fronts increase with increasing the initial difference of dynamic angle of repose.

REFERENCES
[1] OYAMA Y., *Bull. Inst. Phys. Chem. Res. Jpn. Rep.*, **18** (1939) 600.
[2] NAKAGAWA M., ALTOBELLI S. A., CAPRIHAN A., FUKUSHIMA E. *Chem. Eng. Sci.- Shorter Commun.*, **52** (1997) 4423.
[3] HILL K. M., CAPRIHAN A. and KAKALIOS J., *Phys. Rev. Lett.* **78**, 50 (1997).
[4] NEWEY M., OZIK J., VAN DER MEER S.M., OTT E. and LOSERT W. *Europhys. Lett.*, **66**, 205 (2004)

[5] BRIDGWATER J. In *Granular Matter An Interdisciplinary Approach*, Edited By A. Metha (Springer-Verlag, New York) 1994, p. 161

[6] ZIK O., LEVINE D., LIPSON S. G., SHTRIKMAN S. and STAVANS J., *Phys. Rev. Lett.*, **73** (1994) 644.

[7] BOATENG A.A., *Int. J. Multiphase Flow*, **24** (1998) 499.

[8] RISTOW G.H. and NAKAGAWA M., *Phys. Rev. E*, **59** (1999) 2044.

[9] DURY C. M. and RISTOW G. H., *Europhys. Lett.*, **48** (1999) 60.

[10] SANTOMASO A., OLIVI M. and CANU P., *Chem. Eng. Sci.*, **59** (2004) 3269.

[11] DURY C. M., RISTOW G. H., MOSS J. L. and NAKAGAWA M., *Phys. Rev. E*, **57** (1998) 4491.

[12] PARKER D.J., DIJKSTRA A.E., MARTIN T.W. and SEVILLE J.P.K., *Chem. Eng. Sci.*, **52** (1997) 2011.

[13] KOMATSU T.S., INAGAKI S., NAKAGAWA N. and NASUNO S., *Phys. Rev. Lett.*, **86** (2001) 1757.

[14] DURY C. D., RISTOW G. H., MOSS J. L. and NAKAGAWA M., *Phys. Rev. E*, **57** (1998) 4491.